\documentclass[12pt]{article}
\usepackage{graphicx}
\usepackage{nomencl}
\usepackage{amsmath,amsthm,amssymb,bm,array}

\newcommand{\B}{Ba\v zant}  %\newcommand{\BB}{Ba\v zant }  %not needed if I use  \B\ !

     \newcommand{\hhh}{\hspace*{9mm}}

\newcommand{\bc}{\begin{center}}
    \newcommand{\ec}{\end{center}}
\newcommand{\bfr}{\begin{flushright}}
    \newcommand{\efr}{\end{flushright}}
\newcommand{\ii}{\item}
\newcommand{\be}{\begin{enumerate}}
    \newcommand{\ee}{\end{enumerate}}
\newcommand{\bi}{\begin{itemize}}
    \newcommand{\ei}{\end{itemize}}
\newcommand{\bd}{\begin{description}}
    \newcommand{\ed}{\end{description}}
\newcommand{\beq}{\begin{equation}}
\newcommand{\eeq}{\end{equation}}
\newcommand{\bea}{\begin{eqnarray}}
\newcommand{\eea}{\end{eqnarray}}

\newcommand{\bfi}{\begin{figure}}
    \newcommand{\efi}{\end{figure}}
\newcommand{\bay}{\begin{array}{l}}
    \newcommand{\eay}{\end{array}}

\newcommand{\del}{\delta}

\newcommand{\cPr}{\mathbb{P}} %for probability or ...
\newcommand{\cE}{\mathbb{E}} %for probability or ...
\newcommand{\dmg}{\mathcal{D}} % scalar damage

\begin{document}
    \title{\large {\sf
Fishnet Model with Order Statistics for Tail Probability of Failure \\[1mm]
of Nacreous Biomimetic Materials with Softening Interlaminar Links} }
    \author{\sf \large{  Wen Luo\footnote{
Graduate Research Assistant}} and Zden\v ek P. Ba\v zant\footnote{
Corresponding author; McCormick Institute Professor and W.P. Murphy Professor of Civil and Mechanical Engineering and Materials Science, Northwestern University, Evanston, Illinois 60208; z-bazant@northwestern.edu }}
    \maketitle     %\date{  } %5/3 We don't need the date here

    %\tableofcontents

\section{Abstract}
The staggered (or imbricated) lamellar "brick-and-mortar" nanostructure of nacre endows nacre with strength and fracture toughness values exceeding by an order of magnitude those of the constituents, and inspires the advent of new robust biomimetic materials. While many deterministic studies clarified these advantageous features in the mean sense, a closed-form statistical model is indispensable for determining the tail probability of failure in the range of 1 in a million, which is what is demanded for most engineering applications. In the authors' preceding study, the so-called `fishnet' statistics, exemplified by a diagonally pulled fishnet, was conceived to describe the probability distribution. The fishnet links, representing interlaminar bonds, were considered to be elastic perfectly-brittle. However, the links may be quasibrittle or almost ductile, exhibiting gradual postpeak softening in their stress-strain relation. This paper extends the fishnet statistics to links with post-peak softening slope of arbitrary steepness. Probabilistic analysis is enabled by assuming the postpeak softening of a link to occur as a series of finite drops of stress and stiffness. The maximum load of the structure is approximated by the strength of the $k-$th weakest link ($k \geq 1$), and the distribution of structure strength is expressed as a weighted sum of the distributions of order statistics. The analytically obtained probabilities are compared and verified by histograms of strength data obtained by millions of Monte Carlo simulations for each of many nacreous bodies with different link softening steepness and with various overall shapes.

\vskip 2mm
{\bf Key Words:} Failure probability. Fracture mechanics. Structural strength. Monte Carlo simulations. Lamellar structures. Material architecture. Structural safety. Size effect. Scaling. Probability distribution function (pdf). Quasibrittle materials. Brittleness.

\section{Introduction}

The strength and fracture energy of nacre, the shell of pearl oyster or abalone, exceeds by an order-of-magnitude the strength of its constituents (95\% CaCO$_3$). This remarkable property has been shown to originate from the imbricated `brick-and-mortar' arrangement of nanoscale aragonite platelets bonded by a bio-polymer \cite{WangSuo01,Gao03,Shao12,Wei15}.
Thus, the nacre's nanostructure is of great interest for developing new ultra-strong and ultra-tough biomimetic materials.

Most studies have so far been deterministic. However, it is generally agreed that most engineering structures (bridges, airframes, electronic components, etc.) must be designed for failure probability not exceeding $10^{-6}$ per lifetime (which is negligible compared to the risk of death in car accident, $10^{-2}$, and is similar to the risk of death by a falling tree, by lightning, etc.). To design for failure risk $<10^{-6}$, a theoretically based analytical closed-form probability distribution is indispensable. A direct experimental verification of the distribution, by histograms testing, is impossible because $>10^8$ test repetitions would be required to verify the distribution tail at $10^{-6}$, which is obviously beyond reach. Therefore, the experimental verification must rely on other predictions, and the predicted size effect is most useful.

A previous study \cite{LuoBaz17,LuoBaz17PNAS} presented a new statistical model, the fishnet statistics, that can predict the probability tail of a nacreous material in which the fishnet links are perfectly brittle, i.e., their stress drops suddenly to zero as soon as the strength limit of the link is reached. The strength probability distribution of a fishnet was shown to lie between those of a fiber bundle \cite{Dan45,SalBaz14} and of a finite (or infinite) chain \cite{BazLe17, BazLeBaz09, BazPan06, BazPan07, LeBaz09, LeBazBaz11, LeBaz11, BazPla98, Baz05}. These two limiting cases differ in the probability tail by about 2:1, and the fishnet distribution provides a continuous transition between these two limiting cases.

Here we extend the fishnet statistics to links that are quasibrittle, exhibiting progressive postpeak softening of various steepness. The softening may give a more realistic characterization of the interlaminar bond failures in some nacreous structures.

Before reaching the maximum load (which indicates the stability limit and failure if the load is controlled), the fishnet may already contain various numbers, 0, 1, 2,..., of failed links. Based on this observation, the fishnet statistical model splits the fishnet survival event into a union of disjoint events corresponding to different numbers of failed links, which implies a summation of survival probabilities:
 \begin{equation}   \label{fishnet model}
    1-P_f (\sigma) = P_{S_0}(\sigma) + P_{S_1}(\sigma)
                     + P_{S_2}(\sigma) + \cdots
 \end{equation}
where $P_f$ is the failure probability and $P_{S_k}(\sigma)$ ($k=1,2,3,...$) are the probabilities of the whole fishnet surviving under load (or nominal stress) $\sigma$ while there are exactly $k$ failed links under load $\sigma$. This formulation works quite well for fishnets with brittle links but it also poses two difficulties.

First, in obtaining the second and third term ($P_{S_1}$ and $P_{S_2}$) in the foregoing expansion, an equivalent uniform redistributed stress needs to be used for regions near the failed link, based on the stress field from finite element simulation. The error of doing so is negligible when we truncate the expression at the second or third terms (which was shown to give, for brittle links, sufficient accuracy).
But it becomes considerable as more higher-order terms are added. This is because, at the lower tail, $P_{S_k}$ is of the same magnitude as $P_1(\sigma)^k$, so the higher-order terms can easily be ruined by the errors from the previous terms.

Second, as the links become less brittle, more widely scattered damages tend to occur before the peak load, and so more higher-order terms need to be included to predict the failure probability accurately. It is, unfortunately, far more tedious to calculate them. In the previous study, $P_{S_2}$ had to be separated into two parts, to distinguish the cases of two failed links which are either close to, or far away, from each other.

Proceeding similarly, one would have to partition the higher-order terms based on the relative positions of the $k$ failed (or damaged) links and track the stress history for each single case. As $k$ increases, the formula would become too complicated. So the existing fishnet model is suitable only when the links are brittle or almost brittle, in which case it suffices to consider only a few terms in the fishnet expansion (Eq.~(\ref{fishnet model})).

For fishnets with softening links, a modified approach is needed to calculate the failure probability. Although the brittle fishnet model is not applicable to a softening fishnet, its concept has inspired two key ideas to tackle the softening fishnet:
\\ \hhh 1) Instead of a sudden drop of link stiffness to zero stiffness, the progressive continuous postpeak softening is decomposed into a series of sudden stiffness drops, from one link stiffness to the next lower stiffness (Fig.\ref{fig:Model Config}.a).
\\ \hhh 2) The probability distribution of maximum load (or strength) of the softening fishnet is approximated by the order statistics. In other words, for softening links it is no longer the weakest link but the $k$-th weakest link that matters for the maximum load of the whole fishnet. The order statistics allows us to bypass the treatment of complicated stress redistribution in softening fishnets. The order $k$ equals the extent, or number $N_c$, of the link damages right before the maximum load, with $N_c$ being in itself a discrete random variable.

The probability mass function (pmf) of $N_c$ is approximated by the geometric Poisson distribution, which is a distribution appropriate for the random cluster nature of the damages, typical of nacreous systems. This distribution (also called P\'olya-Aeppli distribution) was originally formulated for the insurance industry, to estimate the total cost of claims in a period of time, and later it was used to model the defects in softwares and the process of random word 
substitutions in a DNA molecule \cite{Aep24,Oze10,Ran95,Rob02}.

One common feature of these applications is that they can be described by a collection of clusters in which the number of clusters and the number of objects in each cluster are both random. This feature suggest using 1) the geometric Poisson distribution to model the process of scattered damage accumulation, and 2) the total number of damaged links at the peak load of a softening fishnet.

Combining the distributions of the extent of damage, $N_c$, with the corresponding order statistics, we develop here a probabilistic model for the strength of a fishnet system. To verify the theory, we pick 4 typical softening slopes of links (ranging from almost brittle to almost ductile) and run Monte Carlo simulations $10^6$-times for each case of softening slope.

Finally, it must be stressed that material scientists and engineers developing new materials or structures should strive to maximize not only the mean strength but also the tail strength at probability level  $10^{-6}$. It can happen that a material or structure of a lower mean strength (and the same coefficient of variation) would have a higher tail strength at $10^{-6}$, and vice versa. Cognizant of this fact, we always run at least a million Monte Carlo simulations for each case.

\section{Stochastic Failure: Qualitative Study}

Before embarking on the analytical formulations of failure probability, we begin by presenting some background information and qualitative results, particularly numerical simulations of the stochastic load-displacement curves and damage patterns. This facilitates understanding of the problem to be solved.

\subsection{Numerical Treatment of Softening and Model Configuration}

The numerical method to simulate softening fishnets under uniaxial tension is similar that in \cite{LuoBaz17,LuoBaz17PNAS} where only brittle links are considered. The fishnets are here still treated, in Matlab, as one-dimensional structures, albeit in a collapsed configuration in which the degree of freedom in the transverse direction is omitted. As before, only the strengths of the links are treated as random variables. They obey the following distribution,
    \begin{equation}
        \label{P1 WG}
        P_1(x) =
        \begin{cases}
        2.55 \, \left( 1-e^{-(x/12)^{10}} \right), &
        x \leq 8.6~\text{MPa}\\
        0.526 - 0.474\, \text{erf}[0.884 (10
        - x)] , & x>8.6~\text{MPa}
        \end{cases}
    \end{equation}
The main difference, compared to the previous study, is the numerical realization of softening behavior, which in this study is achieved by replacing continuous softenings with a sequence of small stress drops, or "discrete softenings". This avoids dealing with a tangential stiffness matrix and allows us to use sequentially a linear finite element solver for what is a nonlinear problem.
\begin{figure}[!h]
    \centering
    \includegraphics[width=0.99\textwidth]{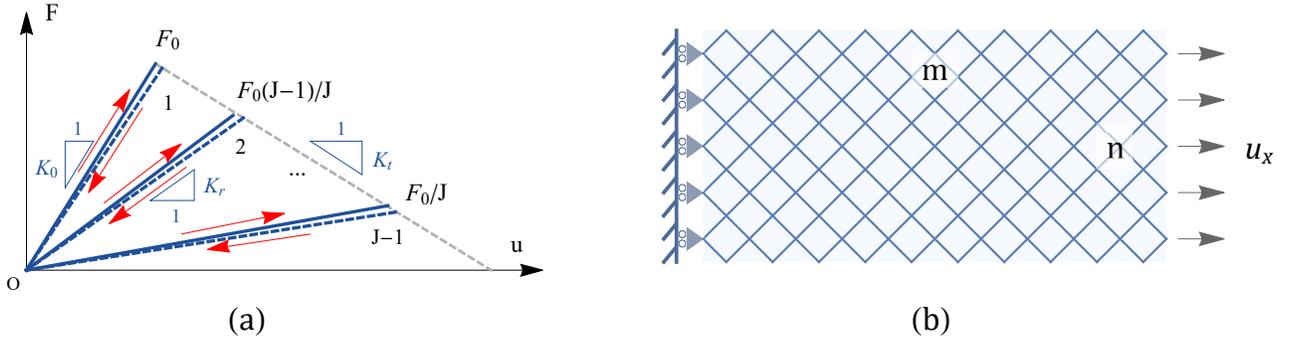}
    \caption{a) Numerical treatment of softening behavior of each single link; b) Schematic showing model configuration and loading conditions.}
    \label{fig:Model Config}
\end{figure}

The idealized constitutive behavior of discrete softening of the links is depicted in Fig.~\ref{fig:Model Config}.a. Each softening link is initially treated as a perfectly elastic-brittle truss element which does not fail immediately after reaching its maximum load ($F_0$) but discontinuously degenerates into, or "jumps to", a weaker linear elastic truss element possessing a slightly lower stiffness and strength. This corresponds to the change of constitutive behavior from $\overline{01}$ to $\overline{02}$. Once the stress in the weakened link has reached its new strength $F_0 (J-1)/J$ (where $J$ denotes the total number of uniform jumps), the link jumps again to the next weaker one. This series of jumps continues until the link strength is finally reduced to 0, in which case the link has fully failed. Clearly, upon increasing $J$, the behavior of the link gradually approaches continuous softening. Here we set $J$ to a relatively large number, such as $J=20$, to reflect a realistic softening while also keeping the computational cost acceptable.

The reduction of link strength after each jump is $F_0/J$. The residual link stiffness follows from the condition that the point $(u_{max},F_{max})$ must lie on the continuous softening curve. Here we consider only linear softening, for which the residual stiffness $K_r$ is a simple function of initial and softening stiffnesses $K_0$ and $K_t$ ($K_t < 0$) and of the damaged softening state $(J-i)/J$, where $i$ is the number of discrete jumps that has already occurred.

The dimension of fishnet is here considered as $N = m \times n = 16 \times 32 = 512$ (see Fig.~\ref{fig:Model Config}.b), length of links $L = 0.01$ mm, cross-section area $A = 1$ $\mathrm{mm^2}$, and Young's modulus $E=1$ MPa. These properties are chosen purely for simplicity since the system is linear elastic. Changing these constants does not change the system's behavior.

As for boundary conditions, the left end of the fishnet is fixed in the x-direction (roller support) while displacement $u_x$ of all the nodes at the other end is prescribed (see Fig.~\ref{fig:Model Config}.b). Thanks to linear elasticity of the links, a simple linear finite element solver could be used to calculate the critical elongation $u_x$ such that, in the current step, one and only one link would be about to soften. So, the variable used to control the numerical process is neither the load nor the displacement but the number of softening jumps, i.e., adding one more discrete softening jump in the structure. After each discrete step, the whole structure is treated as a brand new one and is loaded from the original stress-free state up to the attainment of the strength limit in the the next link, which can be different from the last softened link. A similar procedure was used in the previous simulations of brittle fishnets \cite{LuoBaz17} (and also in previous deterministic simulations of masonry \cite{Gia13}). Thus a detailed algorithm is here omitted.

\subsection{Numerical Results}

To study the behavior of fishnets with various softening tangential stiffnesses, we introduce a fixed random strength field for all simulations. We consider link softening with three subsequent tangential stiffnesses: $K_t = -0.5 K_0$, $K_t = -0.1 K_0$ and $K_t = - 0.01 K_0$ and with two different numbers of softening jumps of the links, $J = 20$ and $J = 500$, while keeping all the other conditions the same.

\subsubsection{Load-Deflection Curves and Stress Evolution}

\begin{figure}[!h]
    \centering
    \includegraphics[width=0.99\textwidth]{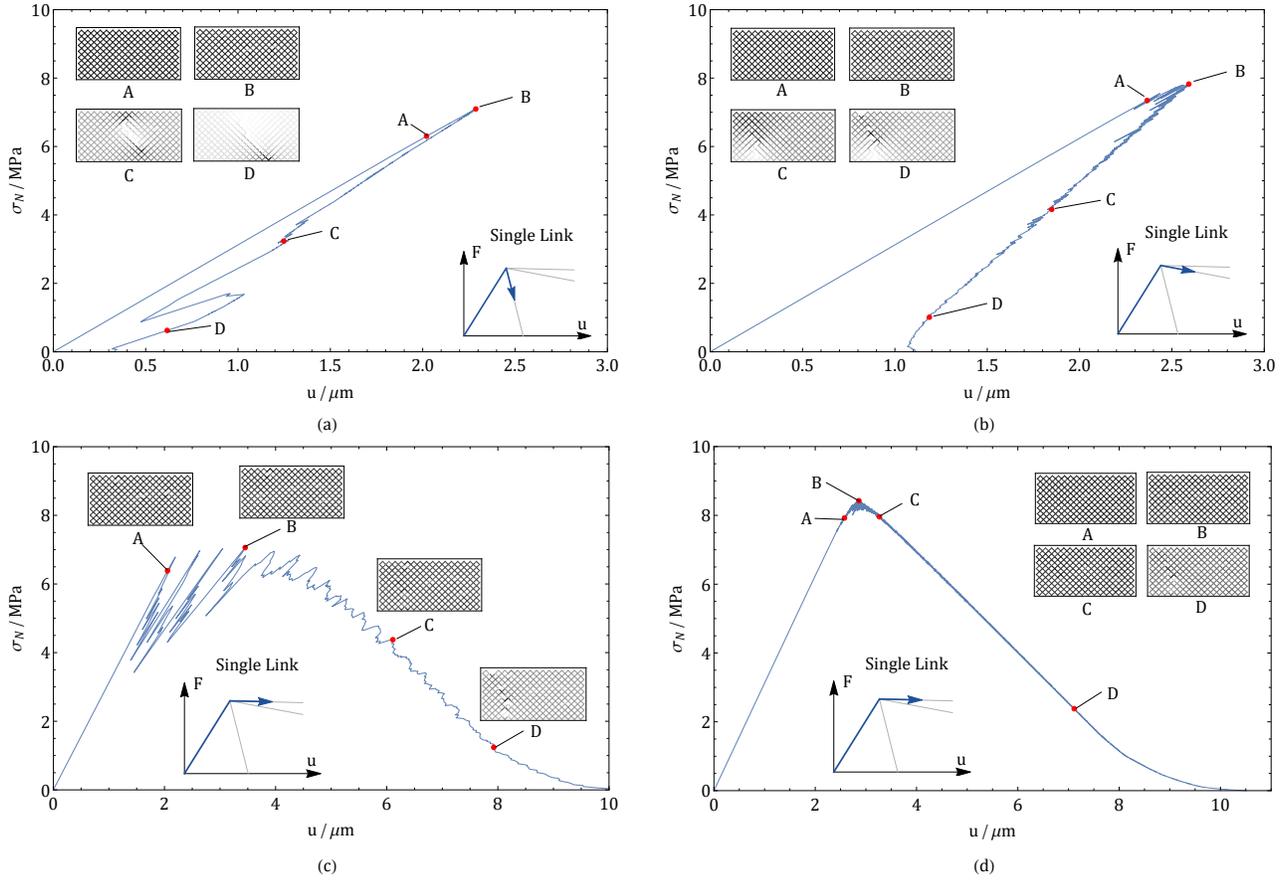}
    \caption{Load-displacement curves for fishnets with the same random link strengths and different tangential softening moduli (and stress fields $\sigma_i/\sigma_{max}$ at 4 typical stages) : a) $K_t = -0.5 K_0$, $J=20$; b) $K_t = -0.1 K_0$, $J=20$; c) $K_t = -0.01 K_0$, $J=20$; d) $K_t = -0.01 K_0$, $J=500$. Magnitude of stress (normalized by the maximum of that frame) is indicated by the darkness.}
\label{fig:Load-Displacement}
\end{figure}

Fig.~\ref{fig:Load-Displacement} shows the load-displacement curves as well as the stress field evolution for various cases. By comparing Fig.\ref{fig:Load-Displacement}a-c, one realizes that the general shape of load-displacement curves (mean behavior) follows the trend shown by the crack band theory \cite{BazOh83} for band fronts of various widths. When $K_t = -0.5 K_0$, the links behavior is close to brittle (i.e., to a vertical stress drop) and the whole fishnet exhibits strong snap-back instability in the post-peak behavior (Fig.~\ref{fig:Load-Displacement}.a) (as already seen in the previous study \cite{LuoBaz17} for brittle links). As the softening slope gets flatter ($K_t = -0.1 K_0$), the snap-back instability is mitigated (Fig.~\ref{fig:Load-Displacement}.b). With an even flatter softening slope for each link ($K_t = -0.01 K_0$), the snap-back instability no longer exists and a post-peak softening curve is observed (Fig.~\ref{fig:Load-Displacement}.c).

Apart from the shape of load-displacement curve, the softening slopes $K_t$ of links are found to have a huge impact on the location of maximum load. As the softening slope of links, $K_t$, increases gradually from $-\infty$ to 0, there is a larger number of softening jumps before the maximum load is reached. Thus the history of stress redistribution gets quite complicated when $K_t$ approaches 0, as can  be seen from the load-displacement curve in Fig.~\ref{fig:Load-Displacement}.c where the maximum load is reached after a long period of wavy stress redistribution. In other words, the problem becomes strongly history dependent when the links are not brittle.

The total number of discrete jumps allowed for each link, $J$, affects the stochastic failure of fishnets as well. Intuitively a larger $J$ would allow a smoother softening process and result in a more realistic stress-strain curve. When $J=20$ (Fig.~\ref{fig:Load-Displacement}.c), a plateau is observed on the load-displacement curve (between point A and B), and it disappears as $J$ is increased to 500 (Fig.~\ref{fig:Load-Displacement}.d). This change of behavior gets more noticeable as the softening slope gets flatter. Intuitively, the reason is that the next discrete softening is more likely to be pushed elsewhere than localize when softening slope is flat, and thus more jumps with smaller stress drops are required to scatter the damaged links. A smaller $J$ imposes larger stress drops at each jump ($F_0/J$) and thus skips many possible ways that could have stopped the damage localization. For  steep softening, though, most damages tend to keep localizing for many steps. Therefore, combining a few consecutive small jumps into a big one does not affect the outcome of softening process, and having a relatively small $J$ is enough to capture the realistic behavior of the nearly brittle fishnet system.

One notable phenomenon in flat softening of the links is that stress field becomes quite smooth and nearly uniform during the whole loading process (Fig.~\ref{fig:Load-Displacement}c and d). This serves as the bases of the mathematical modeling of its failure probability.

\subsubsection{Damage Pattern}
\begin{figure}[!h]
    \centering
    \includegraphics[width=0.9\textwidth]{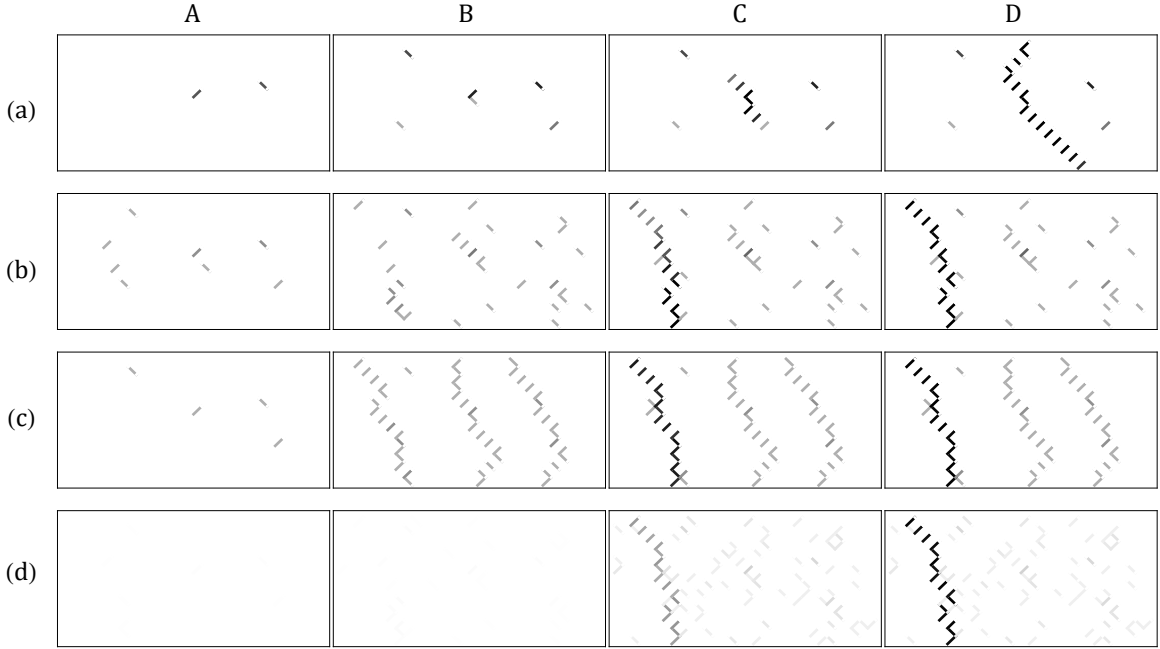}
    \caption{Damage (softening) evolution of the fishnets shown in Fig.~\ref{fig:Load-Displacement}. Rows: (a) $K_t = -0.5 K_0$, $J=20$; (b) $K_t = -0.1 K_0$, $J=20$; (c) $K_t = -0.01 K_0$, $J=20$; (d) $K_t = -0.01 K_0$, $J=500$; Columns: A - prepeak, B - peak, C and D - postpeak. The level of damage $\dmg$ is indicated by the darkness and pure black corresponds to $\dmg=1$.}
\label{fig:Damage Evolution}
\end{figure}

To track the damage evolution during the whole process, we introduce a new variable $\dmg = j/J$, where $j$ is the number of discrete softening jumps that a link has already undergone. $\dmg$ indicates the extent of damage for a single link and $\dmg = 1$ means that the link has failed completely.  Fig.~\ref{fig:Damage Evolution} shows the evolution of $\dmg$ for the cases discussed in the preceding section, in which 4 plots within each row correspond to the damage field at 4 typical stages (A, B, C and D) marked in Fig.~\ref{fig:Load-Displacement}.

Row (a) in Fig.~\ref{fig:Damage Evolution} shows the damage evolution of fishnets with links that are almost brittle (steep softening slope). Before the peak load, only a few scattered damages show up and at each site of damage the damage level has a relatively high value ($\dmg \simeq 0.5$). As the softening slope becomes flatter, more scattered damages show up (row (b), column B), each with a lower value of $\dmg$ compared with case (a) (row (a), column B). In addition, the final crack pattern of case (b) is completely different from case (a) due to the change in the softening behavior of links. For an even flatter softening slope (row (d)-B), the damages at the peak load is too weak to be seen via bare eyes. Note that we do not take case (c) into consideration because, in this case, $J=20$ is not large enough to reflect the realistic softening of links with slope $K_t = - 0.01 K_0$, and thus the three softening bands at the peak load (row(c), column B) are not realistic.

\section{Failure Probability}

In this section, we begin formulating the failure probability of fishnets with softening links. When the softening slope $K_t$ is steep, the behavior of the whole fishnet is almost elastic-brittle. Therefore its failure probability can be very well estimated by the previous fishnet model \cite{LuoBaz17,LuoBaz17PNAS}. However, when $K_t$ is very close to zero, which is more similar to real nacre, the previous model may not be analytically tractable. More specifically, if we want an accurate estimation of $P_f$ we have to consider all possible cases of stress evolution up to the maximum load, which is too tedious to be done for softening fishnet due to its complicated random stress history shown in Fig~\ref{fig:Load-Displacement}.b, c and d. To this end, a different analytical model is needed to accurately predict the failure probability of softening fishnet especially when $K_t$ is nearly 0.

To get the failure probability, we first define a new discrete random variable $N_c$ as the number of previously damaged links upon reaching the peak-load, based on which we then partition the event of structure failure as follows:
 \begin{equation}  \label{general partition}
   P_f(x) = \cPr(\sigma_{max} \leq x)
   =  \sum_{k=0}^{N} \cPr(N_c = k) \cPr(\sigma_{max} \leq x~|~ N_c =k)
 \end{equation}
Then we try to approximate the conditional probability in Eq.~(\ref{general partition}) by the distribution of the $k$-th smallest minimum, $s_{(k)}$, of the strengths of the links:
 \begin{equation}
    \cPr(\sigma_{max} \leq x~|~ N_c = k) \simeq \cPr[s_{(k)} \leq x]
 \end{equation}
Finally we study the distribution of the random variable $N_c$. In a nutshell, the general idea is to approximate $P_f$ using linear combination of the distribution of order statistics (a set of bases) whose combination coefficients are given by the probability $\cPr(N_c=k)$.

\subsection{Bounding Nominal Stress from Above by Order Statistics}

The key part of our model is that we use the $k$-th smallest minimum (i.e., an order statistic) $s_{(k)}$ of the link strengths to bound from above the nominal stress $\sigma_N$ when $N_c=k$. To see why this is possible, we consider the history of stress and residual strength field in time. Note that under uniaxial loading, the nominal stress $\sigma_N$ is defined to be the total load divided by the original cross-section area, which equals the mean value of the redistributed stress field $\sigma_i$. Each fishnet link is assigned an index based on its location (from left to right and top to bottom), so as to identify those that undergo damage.

\begin{figure}[!h]
    \centering
    \includegraphics[width=0.99\textwidth]{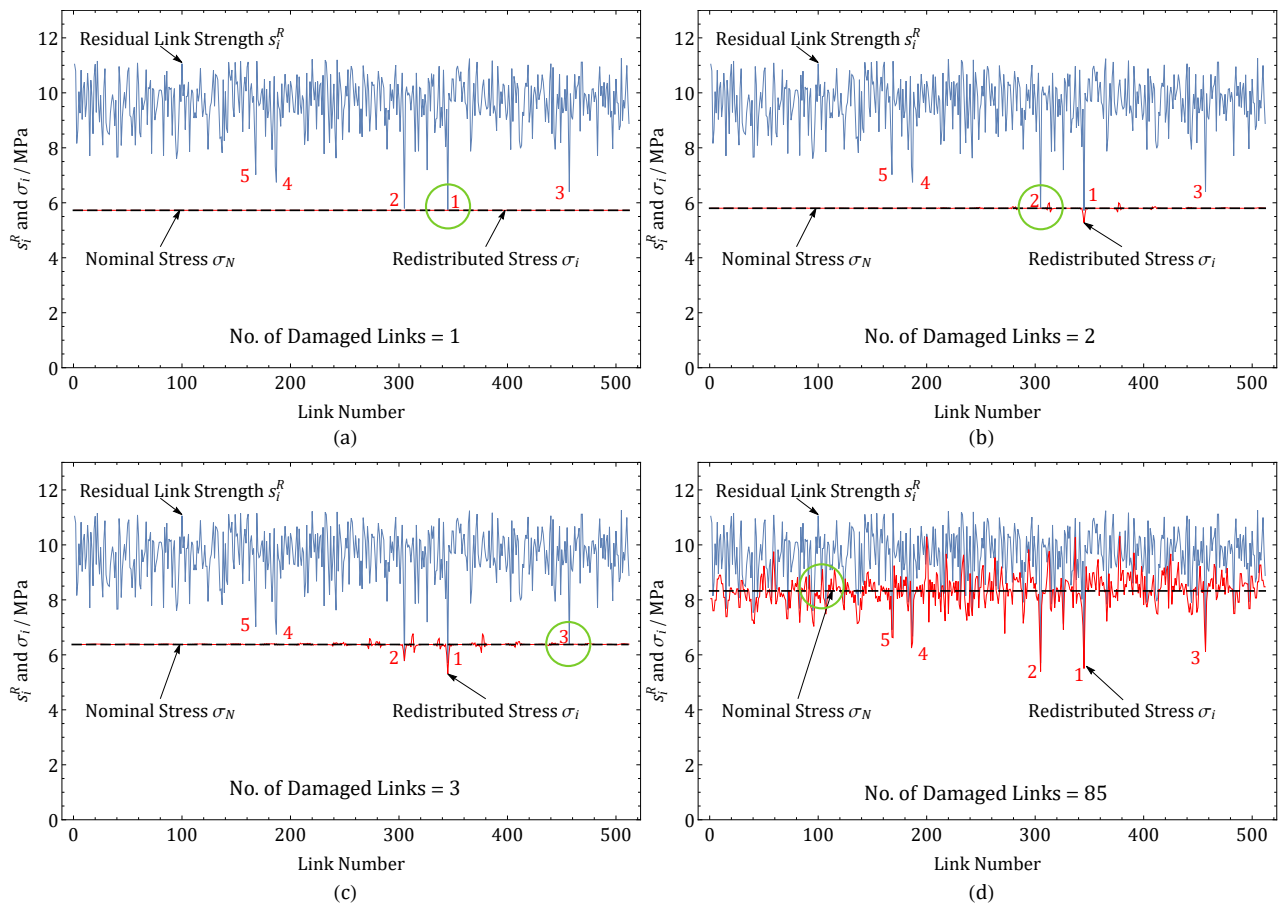}
    \caption{Plot of residual strength field $s_i^R$ and redistributed stress field $\sigma_i$ against link index ($K_t=-0.01K_0$, $J=500$) for a typical random realization. The 5 weakest links are marked with index 1,2,3,4,5 and the circle indicates the link that is to fail at the current step ($s_i^R = \sigma_i$).}
    \label{fig:stress strength evolution}
\end{figure}

Fig.~\ref{fig:stress strength evolution} shows the stress field $\sigma_i$ and residual strength field $s_i^R$ plotted against the link index for four typical stages of a random realization based on the damage extent (measured by the number of damaged links). Note that the 2-D stress and strength fields are plotted as 1-D vectors. In this way, we put aside the geometry temporarily and focus only on the magnitudes of stress and strength for the whole structure. Recall that at each step we use the criterion $s_i^R = \sigma_i$ to find the current softening link, and so the residual strength curve touches the actual stress-strain curve at one and only one place in each frame (marked by the circle) and $s_j^R > \sigma_j$, for all $j \neq i$ is strictly satisfied for the rest of the links, i.e. the rest of links will not undergo further damage under the load in the current step.

Fig.~\ref{fig:stress strength evolution}.a shows the very first step in the numerical simulation, in which the stress field right before the first softening is recorded. The stress field (thin straight line) is strictly constant and equals the strict minimum strength of links $s_{(1)}$, and the weakest link is the first one to undergo softening. Therefore, the assumption that the nominal stress (dashed line) at the $k-$th step equals the strength of the $k-$th weakest link ($ \sigma_N = s_{(1)}$) holds for the first step ($k=1$).

Fig.~\ref{fig:stress strength evolution}.b shows the second step in the simulation. The residual strength field is almost unchanged ($s_{(1)}$ reduced by $s_{(1)}/500$), while some small disturbances in the stress field are observed for a few links due to the decrease of stiffness for the previously softened link. But still the redistributed stress field is almost uniform and most link stresses equal the nominal stress $\sigma_N$ of that step. Most importantly, the stress curve touches the residual strength curve at the second weakest link ($\sigma_N \simeq \sigma_i = s_{(2)}$). So, in the second step, it is no longer the weakest link but the second weakest link that determines the nominal stress and softening process.

Note that, in the second step, the current damage occurred at a different place from the first softening link. If the second damage occurred at the same place as the first one did, or in other words if the damage localizes, the stress at the damaged link will be way below the nominal stress $\sigma_N$ and the equilibrium criterion will be $\sigma_i = 499 s_{(1)}/500 \simeq s_{(1)}$. So, the condition $\sigma_N \simeq s_{(2)}$ will not be satisfied.

Fortunately, damage localization decreases the nominal stress in general and it will increase to a higher level when the localization stops. So the steps where damage localization happens contribute little to the maximum load. We can simply ignore them and only care about the steps in which the damage does not localize. Each time when the damage moves to a new place, the total number of links that are damaged in the fishnet will increase by 1. Therefore we let $k$ denote the total number of damaged links in a fishnet, so as to keep track of the steps where damages did not localize.

\begin{figure}[!h]
    \centering
    \includegraphics[width=0.6\textwidth]{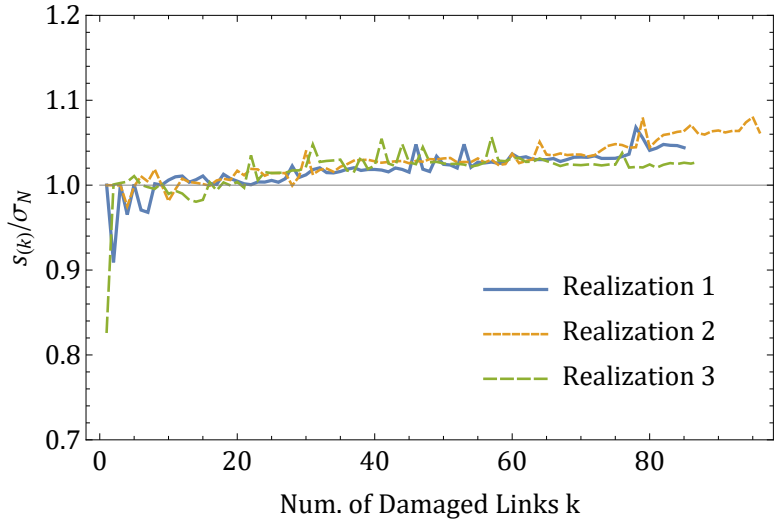}
    \caption{Plot of the quantity $s_k/\sigma_N$ against $k$, number of damaged links, until the maximum load is reached ($K_t = -0.01 K_0$, $J=500$). If there are multiple values of $\sigma_N$ for one $k$, the maximum is used.}
    \label{fig:stress vs strength}
\end{figure}

Fig.~\ref{fig:stress strength evolution}.c shows the fourth step, in which $k=3$ (i.e., the damage localized in the third step and spread out again in the fourth step). Once again we can clearly see that $\sigma_N \simeq s_{(3)}$ holds. As more damages occur, the stress field is no longer nearly uniform when the maximum load is reached (see Fig.\ref{fig:stress strength evolution}.d). To see whether the assumption $\sigma_N \simeq s_{(k)}$ still holds at the maximum load, we plot $s_{(k)}/\sigma_N$ against $k$ up to the step in which the peak load is reached, as shown in Fig.~\ref{fig:stress vs strength}.

Note that if the damages localize in a few steps, $k$ will not increase. Then we choose the largest $\sigma_N$ for various steps with the same $k$. The thick line corresponds to the same realization as used in Fig.~\ref{fig:stress strength evolution}, while the other two dashed lines correspond to two new realizations. One can easily see that, despite the fact that $N_c$ varies a lot for various realizations, the ratio $s_{(k)}/\sigma_N$ varies only little and remains slightly above 1 throughout the process. This justifies our method of using the $k$-th smallest minimum to approximate the nominal stress $\sigma_N$ when number of damaged links equals $k$.

Note that we have so far defined two variables $k$ and $N_c$, which seem to be similar. However, they are not; $N_c$ is a random variable that depends on the maximum load and takes different values for different realizations while $k$ is a deterministic variable that characterizes the number of damaged links at any given time.

\subsection{Relation between nominal stress and order statistics: $\gamma_k$}

The nominal stresses are bounded from above in our formulation ($\sigma_N<s_{(k)}$). If we directly use the distribution of $s_{(k)}$ to replace that of nominal stress $\sigma_N$, we will get a lower bound on the failure probability, which could lead to considerable deviation of our estimate from the true $P_f$. This deviation could be significant when $N_c$ is large, as is the case of flat softening, because the ratio $s_{k}/\sigma_N$ becomes strictly larger than 1 (though only by a small amount) when $k$ is large (see Fig.~\ref{fig:stress vs strength}).

It is therefore necessary to incorporate the qualitative behavior of $s_{k}/\sigma_N$ into our model and approximate $\sigma_N$ by $s_{(k)}/\gamma_k$ instead of $s_{(k)}$ alone, while $\gamma_k$, strictly greater than 1, is taken as the mean curve of $s_{(k)}/\sigma_N$ of a few random realizations up to the maximum load in Fig.~\ref{fig:stress vs strength}. In this way, the distribution of $s_{(k)}/\gamma_k$, compared to that of $s_{(k)}$, is much closer to the distribution of $\sigma_N$.

In fact, if we would not introduce $\gamma_k$, it would lead to a logical contradiction. Suppose that the nominal stress, $\sigma_N$, at the $k-$th step strictly equals $s_{(k)}$. This would give an increasing sequence, and thus the nominal stress would never decrease, contradicting the fact that the maximum load can be reached before the end of displacement control. This observation tells us that $\gamma$ must be considered and that it will always increase to $\infty$ at the end of displacement control ($\sigma_N \simeq s_{(k)}/\gamma_k \rightarrow 0$), even though it is very close to 1 up to the maximum load. Intuitively, $\gamma_k$ indicates the extent of damage localization and stress concentration.

\begin{figure}[!h]
    \centering
    \includegraphics[width=0.99\textwidth]{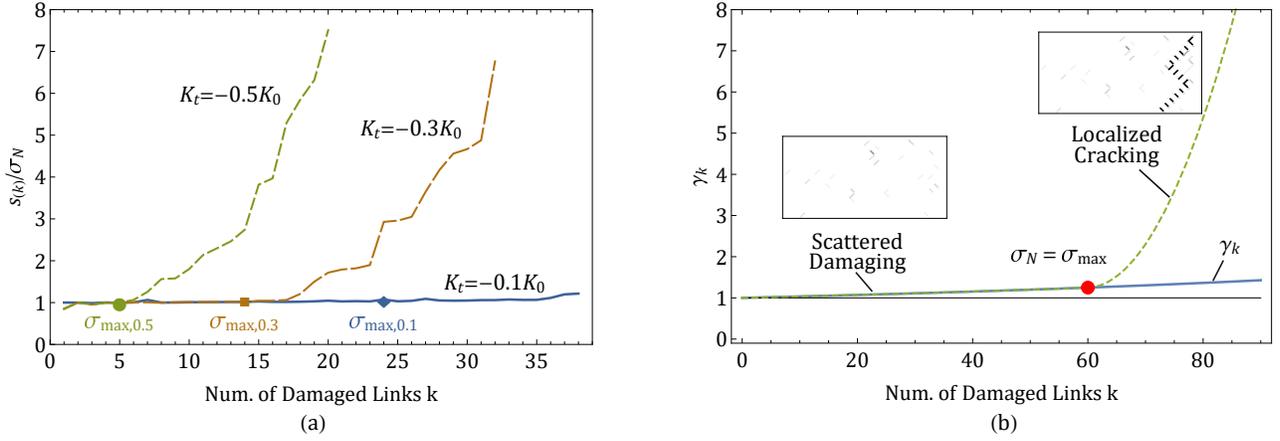}
    \caption{(a) Evolution of the ratio $s_{(k)}/\sigma_N$ against $k$ for various softening slopes $K_t$ ($J=20$); (b) Schematic showing the qualitative behavior of $\gamma_k$.}
    \label{fig:gamma Kt}
\end{figure}

To see the qualitative behavior of $\gamma_k$, we plot in Fig.~\ref{fig:gamma Kt}.a the evolution of the ratio $s_{(k)}/\sigma_N$ for fishnets with 3 different softening slopes ($K_t/K_0=-0.1,-0.3 \text{ and } -0.5$). In all three cases, the same random strength field is used so that the softening slope would be the only variable. It is observed that the curves stay very close to 1 at the beginning. Then, after a critical point, suddenly a sharp transition happens and the curve blows up. The difference, for the flat softening case ($K_t = - 0.1 K_0$), is that $s_{(k)}/\sigma_N$ starts to blow up long after the maximum load while $s_{(k)}/\sigma_N$ tends to blow up much earlier and right after the maximum load, as the softening slope becomes steeper.

This observation matches our conclusion in the previous study \cite{LuoBaz17} of brittle fishnet, namely that maximum load is reached right before damage localization. It is clear that, in all three cases, the maximum load (see the dark disk, square and diamond markers in the figure) is reached before the sharp transition, and so it is guaranteed that our order statistics approximation works.

Note that in Fig.~\ref{fig:gamma Kt}.a, the three curves almost coincide before they blow up and the softening slope $K_t$ affects only the position of the sharp transition. Specifically, the smaller the $|K_t|$ is, the later the sharp transition and the maximum load will materialize. So, $\gamma_k$, defined as the mean curve of $s_{(k)}/\sigma_N$ only up to the maximum load, is independent of the  softening slope $K_t$ although it depends on the fishnet shape and size, and on the number $J$ of the discrete jumps.

Different damage patterns before and after the sudden transition of $s_{(k)}/\sigma_N$ indicate two distinct system behaviors during the whole process---the damages are: 1) small in extent and large in amplitude at the beginning, but 2) large in amplitude and small in extent after the transition (see Fig.~\ref{fig:gamma Kt}.b). Initially, the damages tend to be very weak and uniformly scattered as if they are not too correlated to each other. After the curve bends sharply upward, the damages begin to accumulate and localize in a small region which later forms a contiguous crack. Interestingly, this phenomenon has been noted in \cite{Kra05}, whose authors describe the stochastic damage process of quasi-brittle materials as "ergodic" before the peak load and as of "avalanche class" after the peak load.

\begin{figure}[!h]
    \centering
    \includegraphics[width=0.9\textwidth]{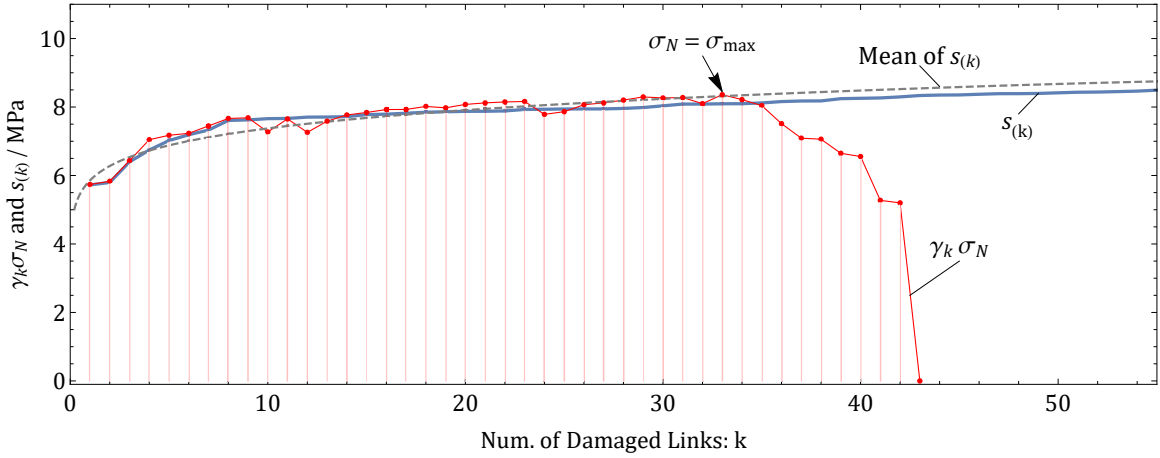}
    \caption{Evolution of scaled nominal stress $\gamma_k \sigma_N$ and order statistics $s_{(k)}$ against $k$}
    \label{fig:sigN vs Sk}
\end{figure}

Fig.~\ref{fig:sigN vs Sk} shows the individual behavior of $\gamma_k \sigma_N$ and $s_{(k)}$ instead of their ratio for a typical random realization ($K_t = -0.2 K_0$). It can be seen that when scattered damages continue showing up, the scaled nominal stress follows closely with the $k-$th smallest minimums until the maximum load is reached, after which damages localize and the nominal stress drastically drops to 0. As a result, the ratio $s_{(k)}/\sigma_N$ blows up after the structure strength gets reduced to zero. It is interesting to note that the general shape of the plot of order statistics $s_{(k)}$ looks similar to $P_1^{-1}(x)$, which can be explained by the inverse transform sampling theory used in pseudo-random number generation.

\subsection{$k$-th Smallest Minima}

According to the foregoing analysis, we approximate the conditional probability in Eq.~(\ref{general partition}) by using the distribution of order statistics (i.e., the $k$-th Smallest Minima):
 \begin{equation}
    \cPr(\sigma_{max} \leq x~|~ N_c = k)
    \simeq \cPr[s_{(k)}/\gamma_k \leq x]
    = \cPr[s_{(k)} \leq \gamma_k x] = W_k(\gamma_kx)
 \end{equation}
where $W_k(x)$ is the cdf of $s_{(k)}$. The detailed derivation of $W_k(x)$ can be found in \cite{Lea12}, and we only give a brief review of order statistics in this section.

The $k$-th smallest minimum is related to the $k$-th largest maximum since
 \begin{equation}
    \min\{X_1,X_2,...,X_n\}  = - \max\{ -X_1,-X_2,...,-X_n \}
 \end{equation}
So it suffices to study the distribution $G_k(x)$ of the $k$-th largest maximum and then the distribution of $s_{(k)}$ can be obtained via the relation:
 \begin{equation}    \label{min max relation}
    W_k(x)=1-G_k(-x)
 \end{equation}

Consider now the exceedances of levels $u_n$ by identical identically distributed (i.i.d.) random variables $X_1, X_2,..., X_n$ having distribution $P_1$. Here $u_n$ is the normalized threshold ($u_n = a_n x + b_n$) used in the extreme value statistics, and $S_n$ denotes the number of exceedances of a level $u_n$ by $X_1,X_2, ..., X_n$. Clearly, the mean of $S_n$ is $n (1 -  P_1(u_n))$, i.e., the total number of random variables times the probability of a random variable being greater than $u_n$. The probability that there are exactly $k$ random variables greater than $u_n$ follows the binomial distribution:
 \begin{equation}
  \cPr\{ S_n \leq k \} = \sum_{s=0}^{k}{n
  \choose s} [1-P_1(u_n)]^{k} P_1(u_n)^{n-k}
 \end{equation}

Now, $n[1-P_1(u_n)] \rightarrow \tau < \infty$ as $n \rightarrow \infty$ represents a condition for $P_1$ to be in the domain of attraction of 1 of the 3 possible types of limiting distributions, So, from the classical Poisson limit for the binomial distribution it follows that
 \begin{equation}
  \cPr\{ S_n \leq k \} \rightarrow e^{-\tau}\sum_{s=0}^{k}
        \frac{\tau^s}{s!}, \text{ as }n \rightarrow \infty
 \end{equation}
where $e^{-\tau}=G_0(x)$ is one of the three types of limiting distributions (Gumbel, Fréchet and Weibull). Therefore,
 \begin{equation}    \label{order max}
  \cPr \{M_n^{(k)} \leq u_n \} = \cPr \{ S_n < k \} = G_0(x)
       \sum_{s=0}^{k}\frac{[-\ln(G_0(x))]^s}{s!}=G_k(x),
 \end{equation}
where $M_n^{(k)}$ is the $k$-th largest maximum of $\{X_1,X_2, ..., X_n\}$. Typically, when $k=0$, $M_n^{(0)}$ is the strict maximum and its distribution follows the classical limit distribution $G_0(x)$. From the relation between $G_k(x)$ and $W_k(x)$, Eq.~(\ref{min max relation}), it follows that
 \begin{equation}         \label{order min}
   W_k(x) = 1 - [1-W_0(x)]\sum_{s=0}^{k}
        \frac{\{-\ln[1-W_0(x)]\}^s}{s!},
 \end{equation}
where $W_0$ is the limiting Weibull distribution.

Alternatively, we could replace $1-W_0(x)$ by $[1-P_1(x)]^N$ if $N$ is large. Note that here we do not replace $x$ by $-x$ because $x$ in Eq.~(\ref{order max}) is negative, representing $-X_i$, while $x$ is positive in Eq.~(\ref{order min}). This means that we have already converted $x$ into its opposite number, and so $x$ represents $X_i$.

\begin{figure}[!h]
    \centering
    \includegraphics[width=0.7\textwidth]{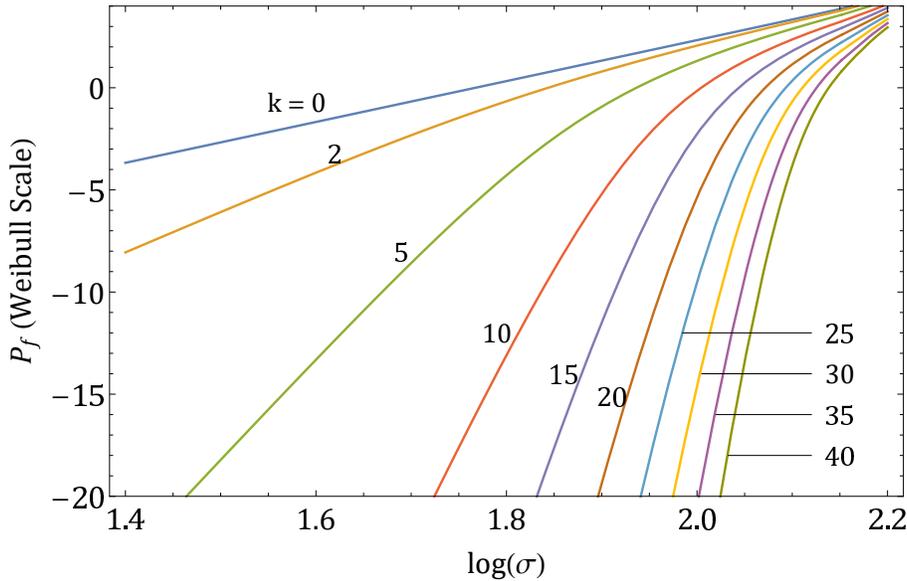}
    \caption{Plot (Weibull scale) of samples from the family of cdf's of the k-th smallest minimum in a collection of 512 i.i.d. random variables that follows the distribution $P_1$.}
    \label{fig:Order Statistics}
\end{figure}

Fig.~\ref{fig:Order Statistics} shows the curves of $W_k(x)$ in Weibull scale for various $k$ values. As one can see, the spectrum of curves spans the region in which the true failure probability $P_f$ could possibly lie and, therefore, it can serve as the basis to represent $P_f$.

\subsection{Critical Number of Softening Links $N_c$}

The failure probability $P_f$ in Eq.~(\ref{general partition}) depends not only on the order statistics but also on the discrete random variable $N_c$, which is the number of damaged links at the peak load. It is again too tedious to get the exact probability mass function (pmf) of the distribution due to its history dependence. We therefore seek an approximation based on the physical nature of $N_c$---the damaged links appear in clusters. To be specific, a cluster keeps growing until the next damage appears far away from the current one, forming a new cluster. Thus, the random variable $N_c$ can be expressed as the following sum,
\begin{equation}
    N_c = \sum_{s=1}^{N_{\text{cluster}}}Y_s,
\end{equation}
where not only the number of damaged links in each cluster $Y_s$ but also the number of clusters $N_{\text{cluster}}$ at maximum load are random variables. We assume that 4$N_{\text{cluster}}$ follows Poisson distribution with parameter $\lambda$ and $Y_s$ follows geometric distribution with parameter $\theta$. Therefore, the sum $N_c$ follows the geometric Poisson (P\'olya-Aeppli) distribution \cite{Aep24}, whose pmf is
\begin{equation}
    p_k = \cPr(N_c = k) =
    \begin{cases}
    \sum_{s=1}^{k}e^{-\lambda}\frac{\lambda}{s!}{k-1 \choose s-1}\theta^s (1-\theta)^{k-s},&~k=1,2,3,\cdots\\
    e^{-\lambda},&~k=0
    \end{cases}
    \label{pmf GeoPoi}
\end{equation}
and the mean and variance are
\begin{equation}
    \cE N_c = \frac{\lambda}{\theta},~~~\mbox{Var}(N_c) = \cE[(N_c-\cE N_c)^2] = \frac{\lambda(2-\theta)}{\theta^2},
    \label{mean and variance of GeoPoi}
\end{equation}
where $\lambda$ is the average number of clusters at maximum load and $\theta$ is the probability of success on each ''trial'', i.e., that the size of current cluster increases by 1.

In our formulation, we assumed that $N_{\text{cluster}}$ follows the Poisson distribution and $Y_s$ follows the geometric distribution. This assumption is true only in the approximate sense, and is the main source of error for the value of $p_k$. More specifically, Poisson distribution of $N_{cluster}$ requires that the clusters would not interact with each other, which is not strictly satisfied because the clusters affect each other through the redistributed stress field. Fortunately, such interactions are very weak, making the assumption justifiable in the approximate sense.

In addition, the geometric distribution of $Y_s$ requires that the probability of success on each "trial" be a constant, which is not strictly satisfied as well. For softening fishnets, the success of a trial can be interpreted as the current softening link lying in the neighborhood of the previous one, so that the size of the current cluster could increase by 1. Due to stress redistribution, the links near the damaged links will have higher stresses than the average, and so the future damages will more likely appear near the damaged links. So, strictly speaking, the probability of success on each trial is not constant.

Nevertheless, in softening fishnets with a rather flat softening slope. the stress concentration is very weak, and then the geometric distribution assumption is applicable in the approximate sense. On the other hand, for relatively brittle fishnets, characterized by steep softening slopes, the geometric Poisson distribution could give noticeable underestimation of $p_k$ for small $k$. To overcome this problem, we will adjust the formulation of $P_f$ by introducing two parameters.

\subsection{Formulation of $P_f$}

Now that we have the distributions of $N_c$ and $s_{(k)}$, the failure probability $P_f$ of softening fishnet in Eq.~(\ref{general partition}) reduces to
 \begin{equation}      \label{failure Prob. Ductile}
    P_f(x) = \sum_{k=0}^{N}p_kW_k\left( \gamma_k x \right),
 \end{equation}
where $p_k$ is given in Eq.~(\ref{pmf GeoPoi}) and $W_k(x)$ by  Eq.~(\ref{order min}). In this formulation, the collection of functions $W_k$ will not change if $P_1$ and $N$ are fixed. So the effect of fishnet shape $m/n$, softening slope $K_t$, and number of allowed jumps for each link $J$ on $P_f$, is embedded in $\gamma_k$ and $p_k$. And since $K_t$ is independent of $\gamma_k$, the softening slope affects the failure probability by influencing the distribution of $N_c$, i.e., the number of damaged links at the peak load.

Note that from Fig.~\ref{fig:stress vs strength}, $\gamma_k$ tends to be, in the first few steps, considerably smaller than 1. So, the first few terms in the sum Eq.~(\ref{failure Prob. Ductile}) could give significant overestimation at the lower tail of $P_f$ when the softening links are not very brittle. Apart from $\gamma_k$, the errors in the estimation of the coefficients $p_k$ by the geometric-Poisson distribution carry over to the final failure probability. To get a more accurate result, we introduce into our model two additional parameters $k_0$ and $\delta k$:
 \begin{equation}    \label{Pf final}
    P_f(x) = \sum_{k=k_0}^{N}p_k W_{k'}\left( \gamma_{k'} x \right),
         ~~~ k' = k-\delta k
    \end{equation}
where $k_0 \geq 1$ and $\delta k \geq 1$ represent, respectively, the truncation and shifting of the terms of order statistics. These two parameters allow us to compensate for a part of the error in the  estimation of $\gamma_k$ and $p_k$. Intuitively, since the difference between the scaled maximum load ($\gamma_k \sigma_{max}$) and order statistics ($s_{(k)}$) is relatively large when the maximum load is reached right after damage initiation, we simply rule out the possibility that the structure could fail when $N_c < k_0$. Hence,  $k_0$ stands for a threshold of truncation for the order statistics. Note that directly omitting the first terms, $k_0$, in Eq.~(\ref{failure Prob. Ductile}) would be incorrect since it would break the partition of unity for coefficients $p_k$, i.e., $\sum_{k=0}^{N}p_k=1$.

So, after the truncation, we always renormalize $p_k$ to ensure satisfying the partition of unity exactly. Apart from the truncation, replacing $k$ by $k-\delta k$ for $W_k(\gamma_k)$ shifts the order statistics to the left in the sequence, which thickens the lower tail of $P_f$. As shown in the following, this adjustment is useful especially when the links are relatively brittle, in which case the geometric Poisson estimation of $p_k$ for small $k$ is not very accurate. On the other hand, when $K_t$ is very flat (i.e, the postpeak almost plastic), $\delta k$ is not needed and can be set to 0. At the same time, since in our model the nominal stresses are not bounded from below, the upper bound on $P_f$ is not guaranteed. Choosing a proper value for $\delta_k$ converts the lower bound on $\del_k$ into an upper bound on the lower tail of $P_f$.

\section{Numerical Verification}

\subsection{Verification of $N_c$}

\begin{figure}[!h]
    \centering
    \includegraphics[width=0.95\textwidth]{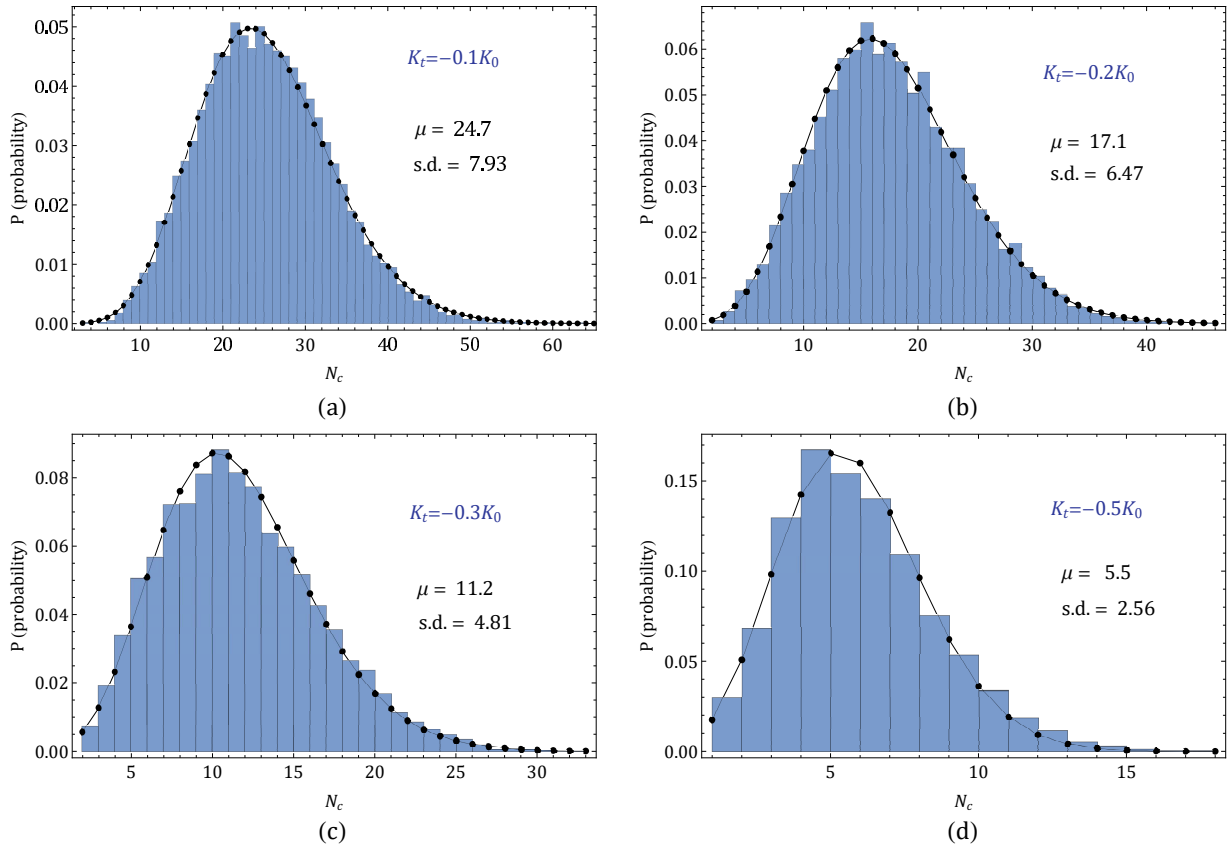}
    \caption{Optimum fit of histograms of $N_c$ with geometric Poisson distributions for fishnets of various softening slopes and same number of discrete link softenings ($J=20$). The mean ($\mu$) and standard deviation ($s.d.$) are shown in the figure.}
    \label{fig:pmf Grid}
\end{figure}

To predict the failure probability accurately, the key is to get an accurate estimation of the coefficients $p_k=\cPr(N_c=k)$, which are assumed to come from the Geometric Poisson distribution. For various softening slopes, $10^4$ Monte Carlo simulations have been run for each case. The histograms as well as the optimum fits of the probability mass functions (pmf) are shown in Fig.~\ref{fig:pmf Grid}. Note that the sample size ($10^4$) chosen here is quite large and could be difficult to achieve through histogram testings in the lab. However, a much smaller sample size, sufficing only for the sample mean and variance, can be used to estimate the two parameters, $\lambda$ and $\theta$, needed for the geometric Poisson distribution (Eq.~(\ref{mean and variance of GeoPoi})).

The histograms from Fig.~\ref{fig:pmf Grid} can be closely fitted, except for some small discrepancies, by the geometric Poisson distribution. As the softening slope becomes steeper, the underestimation by geometric-Poisson distribution becomes larger (Fig.~\ref{fig:pmf Grid}.d) and this could lead to underestimation of $P_f$ at its lower tail. As already mentioned, a shift by $\delta k$ was introduced to partially overcome this problem. Keep in mind, though, that for the limit of steep softening, which is a vertical stress drop, the original fishnet model \cite{LuoBaz17, LuoBaz17PNAS} gives an accurate prediction.

\begin{figure}[!h]
    \centering
    \includegraphics[width=0.6\textwidth]{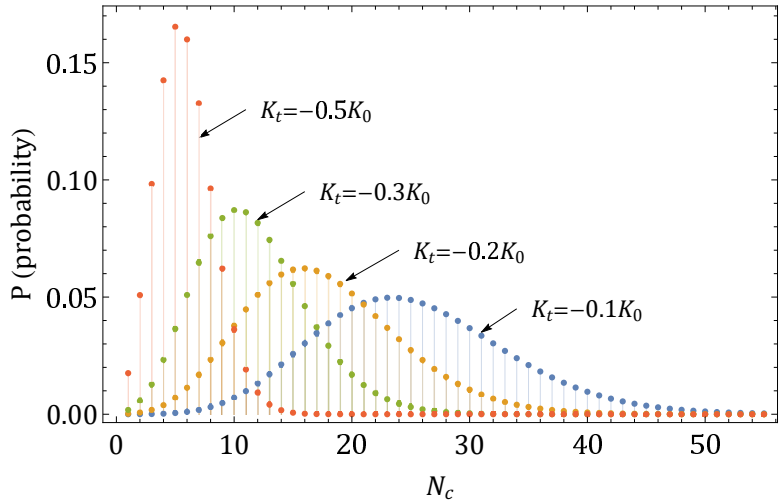}
    \caption{Geometric Poisson distribution of $N_c$ for fishnets of various softening slopes plotted in the same figure.}
    \label{fig:pmf comparison}
\end{figure}

Fig.~\ref{fig:pmf comparison} shows the estimations of the distribution of $N_c$ for various softening slopes, all plotted in one figure. It is clear that, as the softening slope, $K_t$, becomes closer to 0 (quasi-plastic), the mean and variance of $N_c$ increases. This verifies our previous observation that $N_c$ increases as the fishnet links become more plastic (or ductile), while fishnets with brittle links tend to fail at damage initiation. In other words, flatter softening slope delays the occurrence of maximum load.

\subsection{Verification of $P_f$}

\begin{figure}[!h]
    \centering
    \includegraphics[width=0.6\textwidth]{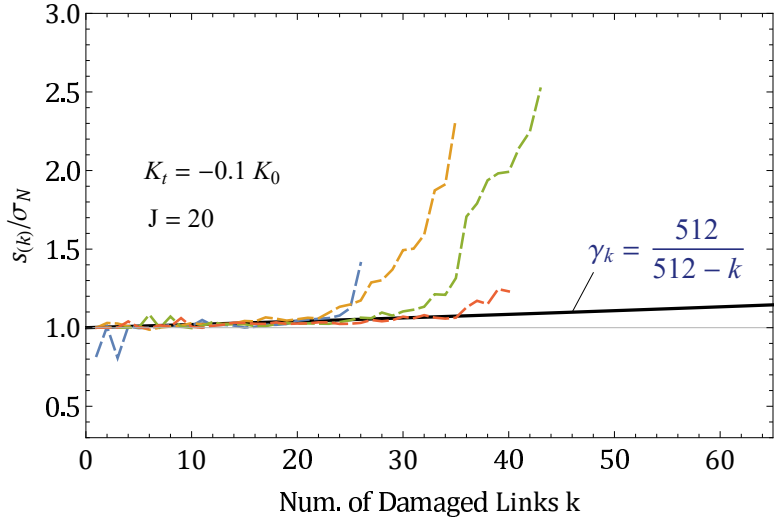}
    \caption{Optimum fit of $\gamma_k$ for the fishnet with $K_t = -0.1 K_0$ and $J = 20$.}
    \label{fig:gamma Fit}
\end{figure}

To describe $P_f$ completely, we must choose an accurate expression for $\gamma_k$. For the current configuration of the numerical model (size, shape and material properties), we let
 \begin{equation}
    \gamma_k = \frac{N}{N-k} = \frac{512}{512-k}
 \end{equation}
This expression is chosen to be the mean curve of $s_{(k)}/\sigma_N$ of a few random realizations before they blow up (see Fig.~\ref{fig:gamma Fit}). Coincidentally, $\gamma_k$ in this particular case is the same as the deterministic expression of $s_{(k)}/\sigma_N$ for a brittle fiber bundle with $N$ fibers.  Alternatively, a linear fit of $\gamma_k$ is also possible since the curve is very flat and smooth.  Once it is chosen, $\gamma_k$ remains fixed for various softening slopes $K_t$ since it only affects the location at which the ratio $s_{(k)}/\sigma_N$ blows up.

\begin{figure}[!h]
    \centering
    \includegraphics[width=0.95\textwidth]{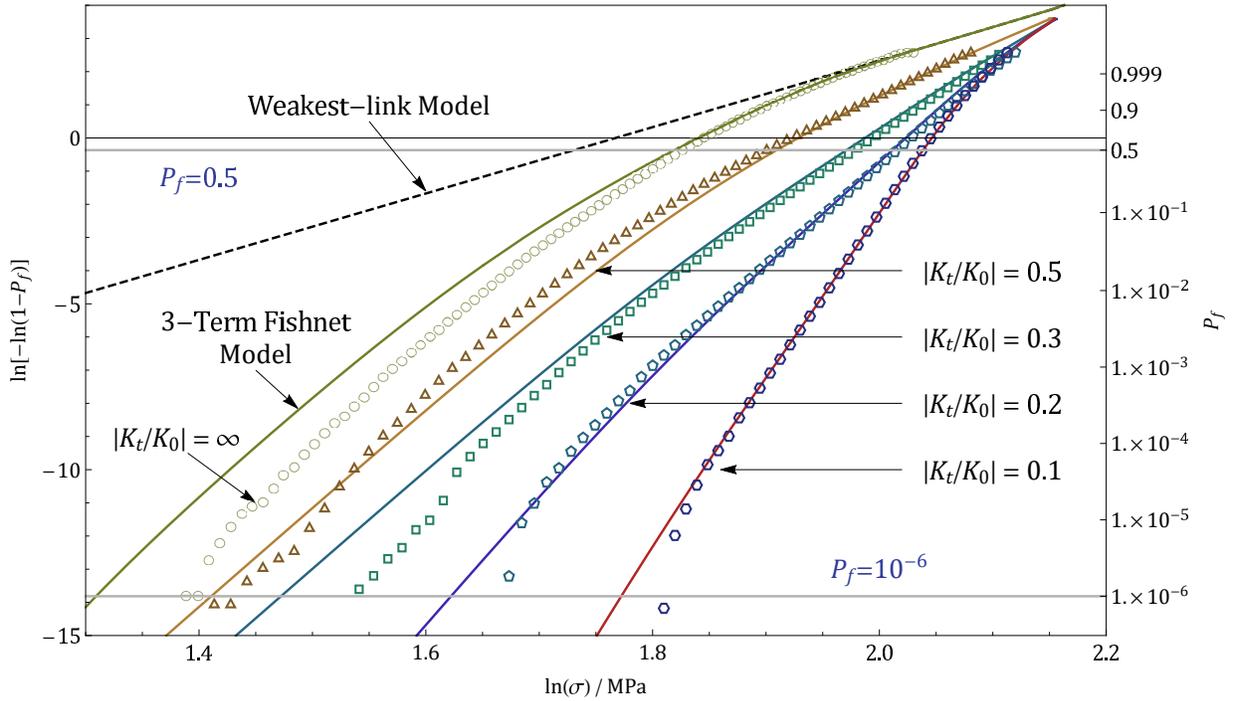}
    \caption{Histogram of nominal strengths (discrete markers) in Weibull scale for fishnets with different softening slopes and results by using order statistics and the original 3-term fishnet model (solid lines). The sample size is $10^6$ for each case.}
    \label{fig:Weibull CDF}
\end{figure}

Fig.~\ref{fig:Weibull CDF} shows the comparison of the failure probability $P_f$ obtained by analytical model and histogram data for various softening slopes. Also included are the histogram and 3-term fishnet model for fishnets with brittle links ($|K_t/K_0|=\infty$) taken from previous study \cite{LuoBaz17}.

\begin{table}[h!]
    \begin{center}
        \caption{parameters $k_0$ and $\delta k$ for cases of various
                            $K_t$}
        \label{tab:table parameters}
        \begin{tabular}{|m{2cm}|m{2cm}|m{2cm}|m{2cm}|m{2cm}|}
            \hline
            $|K_t/K_0|$ & 0.1 & 0.2 & 0.3 & 0.5\\
            \hline
            $k_0$       & 5   & 5   & 5   & 5\\
            \hline
            $\delta k$  & 0   & 2   & 3   & 3\\
            \hline
        \end{tabular}
    \end{center}
\end{table}

Table.~\ref{tab:table parameters} shows the truncation and shifting parameters ($k_0$ and $\delta k$) used in the analytical model. $k_0=5$ is the same for all softening slopes for simplicity and the shifting parameter $\delta k$ gradually increases for increasing magnitude of softening slope $K_t$. It is used to overcome part of the underestimation of $p_k$ and $P_f$ for their lower tails. Note that $\delta k \leq k_0$ should always be satisfied. As to fishnets with very ductile links ($|K_t| \leq 0.1|K_0|$), we set $\delta k=0$ because most fishnets would fail after the size of damage become very large, where the order statistics $s_{k}$ (and their distribution) are close to each other. So shifting the order of $W_k(\gamma_k x)$ makes very little difference and the model without shifting works just fine.

Fig.~\ref{fig:Weibull CDF} shows that the values given by Eq.~(\ref{Pf final}) match quite well the histogram data obtained by Monte Carlo simulations. The discrepancies between the model and data become larger when the softening slope $K_t$ increases. The most accurate estimation is obtained for the quasi-plastic case ($K_t=-0.1K_0$). This is expected because, for quasi-plastic cases, the geometric Poisson distribution of $N_c$ produces very little error. So the model works very well without introducing $k_0$ and $\delta_k$. These  parameters are introduced mainly for the fishnets with links that are quasibrittle to brittle ($|K_t| \gg 0.1$).

Since approximating the distribution of $N_c$ introduces error, one may wonder why not directly use the histogram of $N_c$ for the coefficients $p_k$? Indeed, in this way we could get even better estimations, but the sample size of histogram of $N_c$ in our case is $10^4$, which is still too large for any histogram test to achieve. So in application, using the sample mean and variance to infer the parameters of geometric Poisson distribution remains to be a much more practical and reliable way.

Clearly, the decrease of magnitude of the softening slope $K_t$ makes the whole structure much safer, especially at the $P_f=10^{-6}$ level. When $K_t = -0.5 K_0$, the strength at which $P_f = 10^{-6}$ is about 4.06 MPa, while when $K_t = -0.1 K_0$, this strength increases to about 6.05 MPa---a strength increase of almost 50\%! For comparison, the strength enhancement at the median level ($P_f = 0.5$) is about 22\% (6 MPa to 7.77 MPa). Though the numbers will change for different model configurations, the considerable strength increase at the lower tail of failure probability is found to be a common feature when the softening slope changes from steep to flat.

\section{Conclusions}
        \be \setlength{\itemsep}{-1.5mm}
\ii
In the early stage of uniaxial loading, the damages (or partial stress drops) spawn within the fishnet in a scattered fashion. The nominal stress $\sigma_N$ keeps staying very close to the strength of the $k$-th weakest link $s_{(k)}$, where $k$ is the number of damaged (softened) links in the fishnet.
\ii
After a certain critical moment of loading, new damages cease to be scattered and begin to localize. In the process, the nominal stress $\sigma_N$ begins to deviate from $s_{(k)}$ and quickly drops to 0.
\ii
The softening slope, $K_t$, of links controls the stochastic "time" ($N_c$) of damage localization. A flatter softening slope delays the occurrence of damage localization by increasing the mean of $N_c$.
\ii
Based on the fact that the maximum loads are reached before the damage localization, the structure strength is approximated by the strength of the $N_c$-th weakest link at maximum load $s_{N_c}$, multiplied by a constant, $\gamma_{N_c}$, which is slightly greater than 1. The probability of failure for the whole fishnet can be formulated on the basis of this approximation.
\ii
Both Monte Carlo simulations and analytical results show that fishnets with flatter softening slopes, which may represent the real nacre more closely, have a much higher strength at the extremely low failure probability level, $10^{-6}$, compared with fishnets having steep softening slopes.
\ee

\section{Acknowledgment}

Financial support under ARO Grant W91INF-15-1-0240 to Northwestern University is gratefully acknowledged.

\listoffigures

\end{document}